\documentclass[twocolumn,superscriptaddress,preprintnumbers]{revtex4}
\usepackage{graphicx,bm}
\begin{document}
\newcommand{\mybm}[1]{\mbox{\boldmath$#1$}}
\newcommand{\mysw}[1]{\scriptscriptstyle #1}

\title{Armchair nanographite ribbons and nanotubes turn magnetic: Novel itinerant ferromagnetism without $d$ or $f$ electrons}

\author{Hsiu-Hau Lin}
\email{hsiuhau@phys.nthu.edu.tw}
\affiliation{Department of Physics, National Tsing-Hua University, Hsinchu 300, Taiwan}
\affiliation{Physics Division, National Center for Theoretical Sciences, Hsinchu 300, Taiwan}

\author{Toshiya Hikihara}
\email{hikihara@phys.sci.hokudai.ac.jp}
\affiliation{Division of Physics, Graduate School of Science, Hokkaido University, Sapporo 060-0810, Japan}

\author{Bor-Luen Huang}
\affiliation{Department of Physics, National Tsing-Hua University, Hsinchu 300, Taiwan}

\author{Chung-Yu Mou}
\affiliation{Department of Physics, National Tsing-Hua University, Hsinchu 300, Taiwan}
\affiliation{Physics Division, National Center for Theoretical Sciences, Hsinchu 300, Taiwan}

\author{Xiao Hu}
\affiliation{Computational Materials Science Center, National Institute for Materials Science, Tsukuba 305-0047, Japan}

\date{\today}
\maketitle

\textbf{\textsf{
Merging the functionalities of computation processing and data storage in one single material plays a crucial role in the burgeoning field of spintronics\cite{DasSarma04,Wolf01}. One of the promising candidates is the diluted magnetic semiconductor\cite{MacDonald05,Ohno98} such as (Ga,Mn)As, where itinerant carriers in the semiconducting bands mediate the effective exchange coupling between local magnetic moments of the transition metals and lead to ferromagnetic ground state. Another approach, now growing into a subfield under the name of spin Hall effect, proposes to generate spin currents by electric gates via the spin-orbital interactions and has been verified in experiments.}}

\textbf{\textsf{
Here we would like to propose an alternative non-magnetic resolution, making use of the subtle interplay between Coulomb interactions and lattice topology at nanoscale. By both analytic weak-coupling and numerical density matrix renormalization-group methods, we show that novel itinerant ferromagnetism appears, upon appropriate doping, in (hydrogenated) armchair nanographite ribbons and carbon nanotubes. Inspired by recent experimental breakthrough in single-layer graphite\cite{Novoselov05,Zhang05}, the metallic  and ferromagnetic state we found here may have potential applications in spintronics at nanoscale.}}
\vspace{5mm}

\begin{figure}
\centering
\includegraphics[width=\columnwidth]{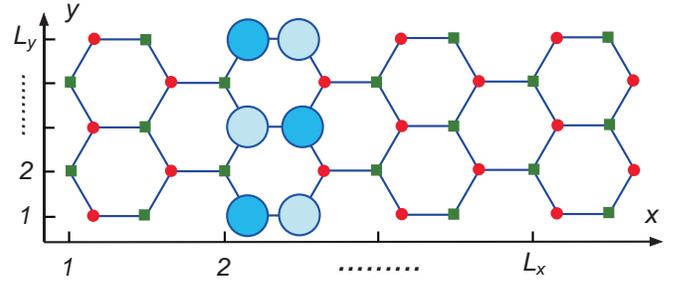}
\caption{\label{fig:Wannier} Armchair NGR of $L_{x}=4$ and $L_{y}=5$. Open edges are present at $y=1$ and $L_{y}$. The solid circles and squares represent sublattice A and B, respectively. The open circles show the amplitudes of the Wannier orbital of $E=t$ at $x=2$, with opposite signs indicated by light/dark shades.}
\end{figure}

In this Letter, we study the ground state properties of the doped armchair nanographite ribbons (NGRs) and carbon nanotubes (CNTs) with short-range interactions. Figure~\ref{fig:Wannier} shows the lattice structure of the armchair NGR with open edges. Even though only $\pi$ electrons are active in low-energy, the honeycomb structure at nanoscale gives rise to both localized Wannier and itinerant Bloch orbitals. Furthermore, we show that these localized orbitals form flat bands with zero velocity. Electron correlations in the flat band generate intrinsic magnetic moments (see below) and the itinerant Bloch electrons mediate ferromagnetic exchange coupling among them. Thus, the magnetic and transport properties can be engineered by changing the width of the NGR and the electric field, opening up the possibility for future spintronics applications at the nanoscale\cite{Meier04,Schneider04,Emberly02,Arita02}.

Besides its possible applications to nanospintronics, the ferromagnetic ground state in the doped armchair NGR also reveals a new type of flat-band ferromagnetism, different from those proposed by Mielke and Tasaki\cite{Mielke93,Tasaki98}. The key to Mielke-Tasaki ferromagnetism is the presence of local Wannier orbitals in the flat band. If the so-called {\em local connectivity condition} for adjacent Wannier orbitals is achieved, the finite overlaps generate exchange coupling among these orbitals, leading to flat-band ferromagnetism. However, the Wannier orbital (shown in Fig.~\ref{fig:Wannier}) in armchair NGR has {\em zero} overlap with its adjacent neighbors. As a consequence, the flat band alone only accounts for the existence of the magnetic moments. As will become clear later, the ferromagnetic order sets in only when additional gapless itinerant carriers are present.

To understand this new type of flat-band ferromagnetism in the armchair NGR, we start with the Hubbard Hamiltonian, $H = H_{t}+H_{U}$,
\begin{eqnarray}
H = -t \sum_{\langle{\bf r,r'}\rangle,\alpha} [c^{\dag}_{\alpha}({\bf r}) c^{}_{\alpha}({\bf r'}) + \mbox{H.c.}]
+ U \sum_{{\bf r}} n_{\uparrow}({\bf r}) n_{\downarrow}({\bf r}),
\end{eqnarray}
where $t$ is the hopping amplitude on the honeycomb network, $U>0$ is the on-site repulsion, $\alpha = \uparrow, \downarrow$ is the spin index, ${\bf r} = (x,y)$, and $\sum_{\langle{\bf r,r'}\rangle}$ is taken only for nearest-neighbor (NN) bonds. Non-uniform hopping due to lattice distortions and short-range interactions with general profiles will be discussed at the end. The values of $t$ reported in the literature\cite{Mintmire92,Wildoer98,Odom98} range from 2.4-2.7 eV for CNTs, while $t \simeq 3$ eV is typical in graphites. Although an accurate value of $U$ is not yet known in nanographite systems, the estimate from polyacetylene, $U \simeq$ 6-10 eV\cite{Baeriswyl85,Jeckelmann94}, might serve as a reasonable guess. Thus, we expect that $U/t \sim {\cal O}(1)$ in nanographite systems. However, to gain physical insight, it is helpful to study the weak coupling limit first.

\begin{figure}
\centering
\includegraphics[width=\columnwidth]{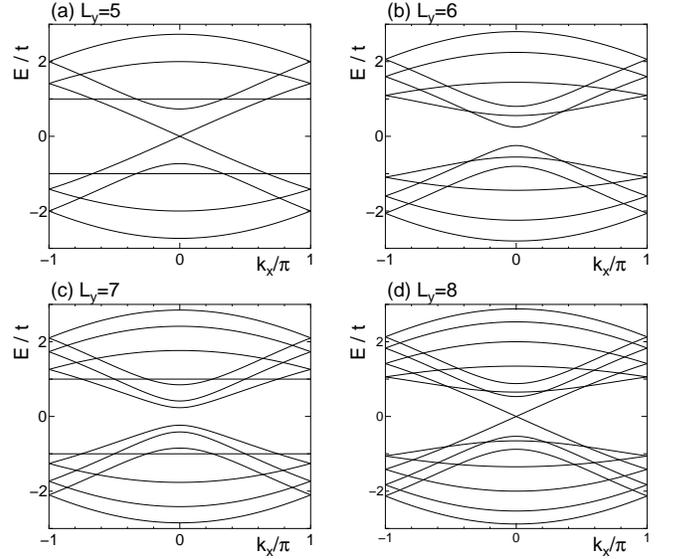}
\caption{\label{fig:Band} Band structure for the infinite armchair NGRs of different width $L_{y}=5,6,7,8$. }
\end{figure}

To simplify the problem, let us take $L_{x} \to \infty$ here. In weak coupling, $U/t \ll 1$, it is natural to concentrate on the hopping Hamiltonian $H_{t}$ first. Its band structure, shown in Fig.~\ref{fig:Band}, can be obtained analytically by partial Fourier transformation and the supersymmetric (SUSY) algebra\cite{Mou04,Lin05}. Note that the flat bands located at $E=\pm t$ appear only for {\em odd} $L_{y}$. The local Wannier orbital at $x=x_{0}$ in the flat band takes the simple form, $\Phi_{\mysw{F}x_0}({\bf r}) = \delta_{x,x_{0}} \Phi_{\mysw{F}}(y)$, where 
\begin{eqnarray}
\Phi_{\mysw{F}}(y) = \left[\begin{array}{cc}
\varphi_{\mysw{A}}(y)\\
\varphi_{\mysw{B}}(y)
\end{array}\right]
=  \frac{1}{\sqrt{L_y+1}} 
\left[
\begin{array}{c}
\sin (\pi y/2)\\
\mp \sin (\pi y/2)
\end{array}\right]
\end{eqnarray}
and the subscripts $A/B$ label the two sublattices in Fig.~\ref{fig:Wannier}. The $\mp$ signs stand for the flat bands at $E=\pm t$. Repeatedly applying the lattice displacement operator $T_{x}$ on one Wannier orbital, all orbitals at different locations can be constructed. Since $[T_{x}, H_{t}] =0$, all the orbitals share the same energy and form a flat band.

At first sight, it is rather counterintuitive that the local Wannier orbital cannot move around by quantum hopping. The static nature is due to perfect destructive quantum interferences which make hopping amplitudes from different sites cancel each other. Furthermore, since the wave function is not zero only at odd $y$ coordinates (see Fig.~\ref{fig:Wannier}), it is clear that the adjacent orbitals have zero overlap. Thus, the Mielke-Tasaki mechanism does not work to couple neighboring orbitals magnetically.

In the following, we concentrate on the flat-band regime $E=t$ for the NGR with odd $L_y$. To account for the low-energy physics correctly, it is necessary to include the dispersive bands intersecting the flat band at Fermi points $k_{m}=2m\pi/(L_{y}+1)$, with $m=1,2,...,\frac12 (L_{y}-1)$. Applying the SUSY technique again, the Bloch wave functions of the itinerant carriers at the Fermi points are obtained as $\Phi_{\mysw{P}m}({\bf r}) = \frac{1}{\sqrt{L_x}}e^{i\mysw{P}k_m x} \Phi_{\mysw{P}m}(y)$, where
\begin{eqnarray}
\Phi_{\mysw{P}m}(y) = \frac{1}{\sqrt{L_y+1}}  \left[
\begin{array}{c}
e^{iP k_{m}\delta_{\mysw{A}}(y)}  \sin (k_m y/2) \\
e^{iP k_{m}\delta_{\mysw{B}}(y)} \sin (k_m y/2)
\end{array}
\right],
\end{eqnarray}
$\delta_{\mysw{A/B}}(y)$ are the phase factors resulting from the honeycomb structure, and $P=R/L=\pm$ denotes the chirality. From the Fermi momenta $k_m$, the lower and upper bounds of the doping rate $x \equiv \langle n \rangle -1$ for the flat-band regime are obtained as $x_{\rm min}=\sum_m k_m/(\pi L_y)=\frac14(1-\frac{1}{L_y})$ and $x_{\rm max}=x_{\rm min} + \frac{1}{L_y}$, respectively.

The effective low-energy theory can be derived from the chiral decomposition of the lattice fermion operator,
\begin{eqnarray}
c_\alpha({\bf r}) \simeq \Phi_{\mysw{F}}(y) \psi_{\mysw{F} \alpha}(x)
+ \sum_{\mysw{P},m} e^{i\mysw{P}k_{m}x} \Phi_{\mysw{P}m}(y) \psi_{\mysw{P}m\alpha}(x),
\label{ChiralDecomposition}
\end{eqnarray}
where $\psi_{\mysw{F} \alpha}(x)$ and $\psi_{\mysw{P}m\alpha}(x)$ are the fermion-field operators for the flat-band and chiral states respectively. Since the density of states is divergent for the flat band, the itinerant chiral fields can be dropped to the leading order approximation . The effective Hamiltonian, keeping the flat band only, is rather simple,
\begin{eqnarray}
H_{\mysw{E}} = U_{\mysw{E}} \sum_{x} n_{\mysw{F}\uparrow}(x) n_{\mysw{F}\downarrow}(x),
\end{eqnarray}
where $U_{\mysw{E}} = U/(L_{y}+1)$ and $n_{\mysw{F}\alpha}(x) = \psi^{\dag}_{\mysw{F}\alpha}(x) \psi^{}_{\mysw{F}\alpha}(x)$. Since the kinetic energy is quenched in the flat band, the ground state contains no quantum fluctuations. To avoid the cost of $U_{\mysw{E}}$, the ground state consists of the maximum number of singly-occupied Wannier orbitals, which leads to local magnetic moments. Since there is no overlap between adjacent orbitals, these magnetic moments are free and give rise to a large ground-state degeneracy.

To lift the large degeneracy of the ground state, interaction with the itinerant carriers {\em must} be included. The effective Hamiltonian can be derived from the complete chiral-field decomposition in Eq.~(\ref{ChiralDecomposition}). The dominant terms in the spin sector are the exchange interactions which couple the local moments and the itinerant carriers,
\begin{eqnarray}
H_{\mysw{J}} =  \sum_{x,m} -J_m\: {\bf S}_{\mysw{F}}(x) \cdot {\bf S}_{m}(x),
\label{ExchangeHamiltonian}
\end{eqnarray}
where ${\bf S}_{\mysw{F}}(x) = \frac{1}{2}\psi_{\mysw{F} \alpha}^\dagger(x) {\bm \sigma}_{\alpha \alpha'} \psi_{\mysw{F} \alpha'}(x)$ is the spin density operator for the local moment (${\bm \sigma}_{\alpha \alpha'}$ is the Pauli matrix) and
${\bf S}_{m}(x) = \sum_{\mysw{P}} {\bf S}_{\mysw{P}m}(x) = \sum_{\mysw{P}} \frac{1}{2}\psi_{\mysw{P}m\alpha}^\dagger(x) {\bm \sigma}_{\alpha \alpha'} \psi_{\mysw{P}m\alpha'}(x)$ for itinerant carriers in each band. The coupling strength is computed from the exchange integral,
\begin{eqnarray}
J_{m} &=& 2 \sum_{{\bf r'}, {\bf r''}} 
\Phi^{*}_{\mysw{F}x}({\bf r'})  \Phi_{\mysw{P}m}({\bf r'})
V({\bf r'}-{\bf r''})
\Phi^{*}_{\mysw{P}m}({\bf r''}) \Phi^{}_{\mysw{F}x}({\bf r''}),
\nonumber\\
&&\hspace{-12mm}= \frac{2U}{(L_y+1)^2}
\sum_{y=odd} 2 \sin^2 (k_m y/2)= \frac{U}{(L_y+1)},
\end{eqnarray}
which is {\em ferromagnetic} for repulsive interaction $U>0$.

The exchange coupling $J_{m}$ tends to align the local moments from the flat band because it does not cause any extra kinetic energy. The ferromagnetically aligned moments act back on the itinerant carriers and induce finite polarization in the itinerant spins. The interacting Hamiltonian $H_{\mysw{E}}+ H_{\mysw{J}}$ therefore shows interesting two-step flat-band ferromagnetism -- the flat band gives rise to local moments without direct exchange coupling, while the presence of gapless itinerant carriers mediates the ferromagnetic order. The significant feature of the armchair NGR/CNT is that, even within the one-orbital Hubbard model without any magnetic impurity nor additional localized levels, the electronic correlations give rise to {\em both} the local moments and itinerant carriers simultaneously due to the peculiar topology of the honeycomb lattice.

Now we turn to armchair NGRs with finite $L_{x}$. If one imposes the periodic boundary conditions along the $x$-axis, the system becomes a short segment of armchair CNT. In this case, only when the quantized momenta $k_{x} = 2n\pi/L_{x}$ coincide with the Fermi points $k_{m} = \pm 2\pi m/(L_{y}+1)$, the gapless itinerant carriers are present and the ferromagnetic ground state is realized. For open boundary conditions with finite $L_x$, the situation is much more complicated; $k_{x}$ is no longer good quantum number and the band structure can be deformed by finite $L_x$. However, from numerical calculations for the tight-binding Hamiltonian $H_{t}$, we have found that the energy spectrum around the flat-band level $E = \pm t$ is not affected by finite $L_x$ and can be accurately approximated by the quantization rule $k_{x} = l\pi/(L_{x}+1)$ ($l = 1, 2, ..., L_x$). When the quantized momenta $k_{x}$ coincide with the Fermi points $k_{m}$, ferromagnetism sets in with the help of these gapless itinerant carriers. Therefore, depending on specific choice of $L_{x}$, the ground state of the armchair NGR/CNT in the flat-band regime can be ferromagnetic or Curie-like paramagnetic. 

\begin{figure}
\centering
\includegraphics[width=\columnwidth]{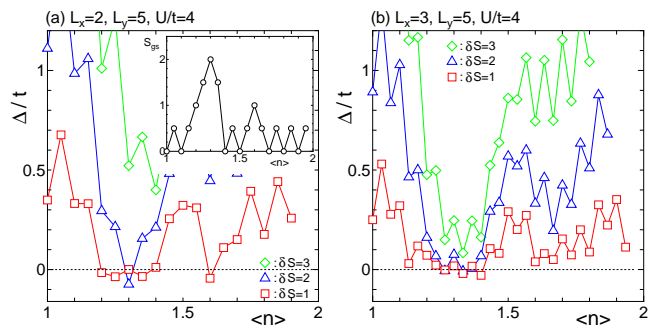}
\caption{\label{fig:GAP} Doping-rate dependence of the energy differences between higher-spin and lowest-spin states in the $L_y = 5$ armchair NGRs with (a) $L_x = 2$ and (b) $L_x = 3$. The squares, triangles, and diamonds represent the data for $\delta S = 1$, $2$, and $3$, respectively. Inset in (a): Total spin of the ground state, $S_{\rm gs}$, for $L_{x}=2$ and $L_{y}=5$ armchair NGR at different doping.}
\end{figure}

\begin{figure}
\centering
\includegraphics[width=\columnwidth]{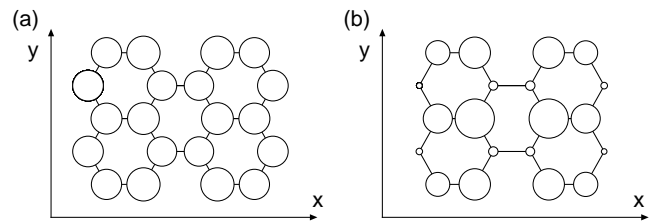}
\caption{\label{fig:WF2} Ground-state distribution of (a) charge density $\langle n({\bf r}) \rangle -1$ and (b) spin polarization $\langle S^z({\bf r}) \rangle$ for $L_{x}=2$ and $L_{y}=5$ armchair NGR with $(N_{e} = 5, \delta S=1)$. The values of $\langle n({\bf r}) \rangle -1$ and $\langle S^z({\bf r}) \rangle$ are positive at every site, and the areas of the open circles are proportional to their amplitudes.}
\end{figure}

To check whether the weak-coupling picture survives for the realistic coupling
regime $U/t \sim {\cal O}(1)$, we employ the density-matrix renormalization
group (DMRG) method\cite{White92} to study the system in intermediate-coupling regime
$U/t=4$. To look for higher-spin ground state, we perform the calculation
in a subspace preserving total spin of the system by using the SU(2)-symmetric
density matrix. We compute the energy differences between
the higher-spin ground states and lowest-spin ones defined as,
\begin{eqnarray}
\Delta(N_{e}, \delta S) = E_{0}(N_{e}, S_{0}+\delta S) - E_{0}(N_{e},S_{0}),
\end{eqnarray}
where $N_{e}$ denoting the number of doped electrons from half filling $\langle n \rangle =1$ and $E_{0}(N_{e}, S)$ is the lowest energy in the subspace characterized by $N_e$ and total spin $S$. The lowest spin $S_{0}$ is defined as $S_{0}=0$ for even $N_{e}$ and $S_{0}=1/2$ for odd $N_{e}$. The calculation is performed for the system with $L_y = 5$ and $L_x = 2, 3$. The number of kept states is up to $1200$, typically corresponding to $400$-$500$ SU(2) multiplets. The truncation error is typically $10^{-5}$ for $L_x = 2$ and $10^{-4}$ for $L_x = 3$, and the results are extrapolated to the limit of zero truncation error.

For $L_{x}=2$, the quantized momenta $k_x = l \pi/3$ coincide with the Fermi points of the chiral fields, and therefore, the weak-coupling theory predicts the ferromagnetic ground state. As shown in Fig.~\ref{fig:GAP}, we do find numerically the ferromagnetic (higher-spin) ground state in the flat-band regime ($N_{e}=5,6,7$). The numerical values of $\Delta$ for these higher-spin ground states are $\Delta(5, \delta S=1) = -0.036t$, $\Delta(6, \delta S=2) = -0.074t$ and $\Delta(7, \delta S=1) = -0.035t$, respectively. In addition, the local charge density $\langle n({\bf r}) \rangle -1$ and spin polarization $\langle S^z({\bf r}) \rangle$ in the ground state are also calculated, shown in Fig.~\ref{fig:WF2}. While the charge distribution is almost uniform, it is rather remarkable that the spin density has a similar profile to the Wannier orbitals in the weak-coupling limit. This indicates that the physical picture developed in weak coupling still works rather well even for $U/t=4$.

For $L_{x}=3$, on the other hand, the approximative quantization rule gives $k_{x} = l\pi/4$, so that no gapless itinerant carriers in weak-coupling theory are available. In the numerical calculation, we find that in the flat-band regime, $N_{e} = 7,8,9,10,$ and $11$, the higher-spin ground state only shows up for $N_{e}=8, 10$. The numerical values are $\Delta(8, \delta S=2) = -0.006t$, $\Delta(10, \delta S=1) = -0.02t$ -- relatively smaller than those in $L_{x}=2$ case. Since the energy differences are very small, it is quite natural to consider that the lowest- and higher-spin states are almost degenerate, suggesting a paramagnetic ground state with almost free moments. We therefore believe that the numerical results support the novel finite-size effect predicted by the weak-coupling analysis.

Before closing, it is interesting to ask how robust the flat-band ferromagnetism is beyond the Hubbard Hamiltonian. Note that the flat band in armchair NGR is unique in several ways. First of all, to achieve flat bands in other systems, one needs to fine-tune several parameters while it is not necessary for armchair NGR -- when the transverse width $L_{y}$ is odd, the presence of the flat bands is guaranteed by the topology of the lattice structure. In addition, in practical graphite network, the hopping along horizontal and titled bonds is expected to be slightly different. It is straightforward to show that this deviation only shifts the flat band from $E=\pm t$ but the bands remain flat. The profile of the short-range interaction does not seem to do much harm either. In the presence of the NN interaction $V$, the exchange coupling in Eq.~(\ref{ExchangeHamiltonian}) becomes
\begin{eqnarray}
J_{m} = \frac{1}{(L_y+1)} (U-V \cos k_m).
\end{eqnarray}
Since $V<U$ is often expected, the exchange coupling is still ferromagnetic and the picture does not change.

Recently NGRs have been successfully fabricated\cite{Kyotani03} and may bring up the possibility to realize this new kind of flat-band ferromagnetism. Meanwhile, the developing nanotechnology opens up another window to probe the interesting phenomena in a nanojunction by crossing a semiconducting CNT on top of a single-wall armchair one. Applying gate voltages in the semiconducting nanotube to deplete electrons locally, it is possible to produce the desired spatial profile of the density so that some regimes are in the flat-band ferromagnetic/paramagnetic phase. The key issue is whether it is feasible to achieve sharp density profile by gate voltage alone. The most promising approach is provided by recent breakthrough in the fabrication of single-layer graphite\cite{Novoselov05,Zhang05} with high mobility. Through the four-probe transport measurement, it was demonstrated that the carrier density can be controlled at will by electric gates. Therefore, if one can shrink the width of the single-layer graphite, the signature of the novel ferromagnetism at nanoscale will start to emerge.

In summary, combining the weak-coupling analysis and the DMRG technique, we have shown that the armchair NGR, due to electronic correlations and the flat band, has intrinsic magnetic moments. Furthermore, if gapless itinerant carriers are present, they mediate exchange coupling among the moments and lead to the ferromagnetic ground state.

We acknowledge Leon Balents, Greg Fiete, Yukitoshi Motome and Tsutomu Momoi for valuable discussions and comments. HHL, BLH and CYM appreciates financial supports from National Science Council in Taiwan. The hospitality of KITP are also greatly appreciated.

\end{document}